\documentclass[floatfix,aps,prl,twocolumn,showpacs,superscriptaddress]{revtex4-1}
\usepackage{float}
\usepackage[dvipsnames]{xcolor}
\usepackage{graphicx}
\usepackage{ulem}
\usepackage{epstopdf}
\usepackage{amsfonts}
\usepackage{amsmath}
\usepackage{mathtools}
\usepackage{bbold}
\usepackage{booktabs}
\usepackage{cases}
\usepackage{siunitx}
\usepackage[toc,page]{appendix}

\usepackage{tikz}
\usetikzlibrary{positioning}
\usetikzlibrary{backgrounds}
\usetikzlibrary{shapes,calc,patterns,
	decorations.pathmorphing,
	decorations.markings}

\pgfmathdeclarefunction{gauss}{3}{%
	\pgfmathparse{exp(-((#1-#2)^2)/(2*#3^2))}
}

\pgfmathdeclarefunction{gaussmath}{3}{%
	\pgfmathparse{1/(#3*sqrt(2*pi))*exp(-((#1-#2)^2)/(2*#3^2))}
}
	
\definecolor{cargoblue}{rgb}{0.145, 0.204, 0.58}
\definecolor{parAgreen}{rgb}{0.137, 0.545,0.271}
\definecolor{springred}{rgb}{0.843, 0.188, 0.122}
\definecolor{springred2}{rgb}{0.843, 0.188, 0.122}
\definecolor{maroon}{rgb}{0.502,0,0}

\definecolor{green-light}{rgb}{0.678,0.867,0.557}
\definecolor{green-medium}{rgb}{0.255,0.671,0.365}
\definecolor{green-dark}{rgb}{0,0.353,0.196}

\definecolor{red-light}{rgb}{0.992,0.733,0.518}
\definecolor{red-medium}{rgb}{0.937,0.396,0.282}
\definecolor{red-dark}{rgb}{0.60,0,0}

\definecolor{blue-light}{rgb}{0.5, 0.80, 0.73}
\definecolor{blue-medium}{rgb}{0.11, 0.57, 0.75}
\definecolor{blue-dark}{rgb}{0.047,0.173,0.518}

\newcommand\chout{\bgroup\markoverwith{\textcolor{red}{\rule[0.5ex]{2pt}{1.0pt}}}\ULon}

\usepackage{soul}

\begin{document}
\title{Theory of Active Intracellular Transport by DNA-relaying}

\author{Christian Hanauer}
\thanks{These authors contributed equally to this work.}
\affiliation{Arnold-Sommerfeld-Center for Theoretical Physics and Center for
	NanoScience, Ludwig-Maximilians-Universit\"at M\"unchen,
	D-80333 M\"unchen, Germany.}

\author{Silke Bergeler}
\thanks{These authors contributed equally to this work.}
\affiliation{Arnold-Sommerfeld-Center for Theoretical Physics and Center for
	NanoScience, Ludwig-Maximilians-Universit\"at M\"unchen,
	D-80333 M\"unchen, Germany.}

\author{Erwin Frey}
\affiliation{Arnold-Sommerfeld-Center for Theoretical Physics and Center for
	NanoScience, Ludwig-Maximilians-Universit\"at M\"unchen,
	D-80333 M\"unchen, Germany.}

\author{Chase P. Broedersz}
\email{c.p.broedersz@vu.nl}
\affiliation{Arnold-Sommerfeld-Center for Theoretical Physics and Center for
	NanoScience, Ludwig-Maximilians-Universit\"at M\"unchen,
	D-80333 M\"unchen, Germany.}
\affiliation{Department of Physics and Astronomy, Vrije Universiteit Amsterdam, 1081 HV Amsterdam, The Netherlands}

\date{\today}

\begin{abstract}	
The spatiotemporal organization of bacterial cells is crucial for the active segregation of replicating chromosomes. 
In several species, including \textit{Caulobacter crescentus}, the ATPase ParA binds to DNA and forms a gradient along the long cell axis. 
The ParB partitioning complex on the newly replicated chromosome translocates up this ParA gradient, thereby contributing to chromosome segregation.  
A DNA-relay mechanism---deriving from the elasticity of the fluctuating chromosome---has been proposed as the driving force for this cargo translocation, but a mechanistic theoretical description remains elusive. Here, we propose a minimal model to describe force generation by the DNA-relay mechanism over a broad range of operational conditions.
Conceptually, we identify four distinct force-generation regimes characterized by their dependence on chromosome fluctuations. These relay force regimes arise from an interplay of the imposed ParA gradient, chromosome fluctuations, and an emergent friction force due chromosome-cargo interactions.
\end{abstract}

\maketitle
\noindent
The interior organization of bacterial cells is an essential prerequisite for several vital processes, ranging from chromosome and plasmid segregation to cell division~\cite{Surovtsev2018}.
Dedicated active mechanisms ensure the rapid translocation and accurate localization of macromolecular objects, such as low-copy-number plasmids~\cite{Toro2010}, protein clusters~\cite{Schumacher2017a,Roberts2012}, and carboxysomes~\cite{Savage2010}. 
A prominent example is the translocation of the partition complex in bacteria such as \textit{Caulobacter crescentus}. 
One copy of the partition complex --- bound to the newly replicated chromosome --- translocates rapidly from the old to the new cell pole, resulting in chromosome segregation~\cite{Jensen1999}.  
The translocation of the chromosome-bound partition complex depends on a protein gradient: the partition complex follows an increasing amount of the ATPase ParA in the cell~\cite{Ptacin2010, Schofield2010, Shebelut2010, Surovtsev2016b}.
However, the physical principles underlying this directed motion of the partition complex remain unclear.

The ATPase ParA belongs to the widely conserved ParAB\textit{S} partitioning system for chromosome and plasmid segregation~\cite{Livny2007}. 
The partitioning complex is a large centromere-like protein-DNA cluster consisting of interacting ParB proteins~\cite{Livny2007,Broedersz2014,Mohl1997,Murray2006,Breier2007}.
The ATPase ParA exists in an ADP- and ATP-bound form and its prefered location in the cell can change dependent on its nucleotide state~\cite{Lutkenhaus2012}:
As an ATP-bound dimer, ParA binds nonspecifically to DNA and, upon interaction with ParB, its ATPase activity is stimulated leading to detachment of ADP-bound ParA monomers into the cytosol. 
The interactions of ParA ATPases with the partition complex are necessary for its directed translocation~\cite{Lutkenhaus2012}. 

Various mechanisms have been proposed for force generation~\cite{Ringgaard2009, Ptacin2010, Banigan2011, Shtylla2012,Sugawara2011}, including a class of Brownian-ratchet models~\cite{Lim2014,Hu2015, Hu2017a, Hu2017, Walter2017}. Specifically, a DNA-relay mechanism was suggested~\cite{Lim2014,Wiggins2010}, where DNA-bound ParA proteins relay the partition complex up a ParA concentration gradient by exploiting elastic fluctuations of the chromosome~\cite{Lim2014,Surovtsev2016a}.
It has been argued using simulations, that this model can explain the experimentally observed translocation of the partition complex~\cite{Lim2014, Surovtsev2016a}. However, a theoretical description of the DNA-relay force that reveals the dependence of the force on key system parameters is still lacking. 

Here, we present an analytic theory for force generation by the DNA-relay mechanism. 
We compute the relay force by evaluating the stochastic binding of DNA-bound ParA-like proteins to a cargo using a Master equation approach. 
Conceptually, the predicted relay force originates from the interplay of the ParA gradient, chromosome fluctuations, and an emergent friction force due to the interactions of chromosome-bound ParA proteins with the cargo. 
These contributions give rise to four distinct force generation regimes, depending on the strength of chromosomal fluctuations and the cytoplasmic friction on the cargo. 
We thus establish a theoretical framework to characterize the DNA-relay mechanism over a broad range of operational conditions, providing conceptual insight into active directed transport of ParB-like cargos for \textit{in vivo} \cite{Lim2014,Ietswaart2014,Schumacher2017,LeGall2016} and \textit{in vitro} \cite{Vecchiarelli2014} settings.     

\begin{figure}[t!]
	\centering
	
	\includegraphics[width=\columnwidth]{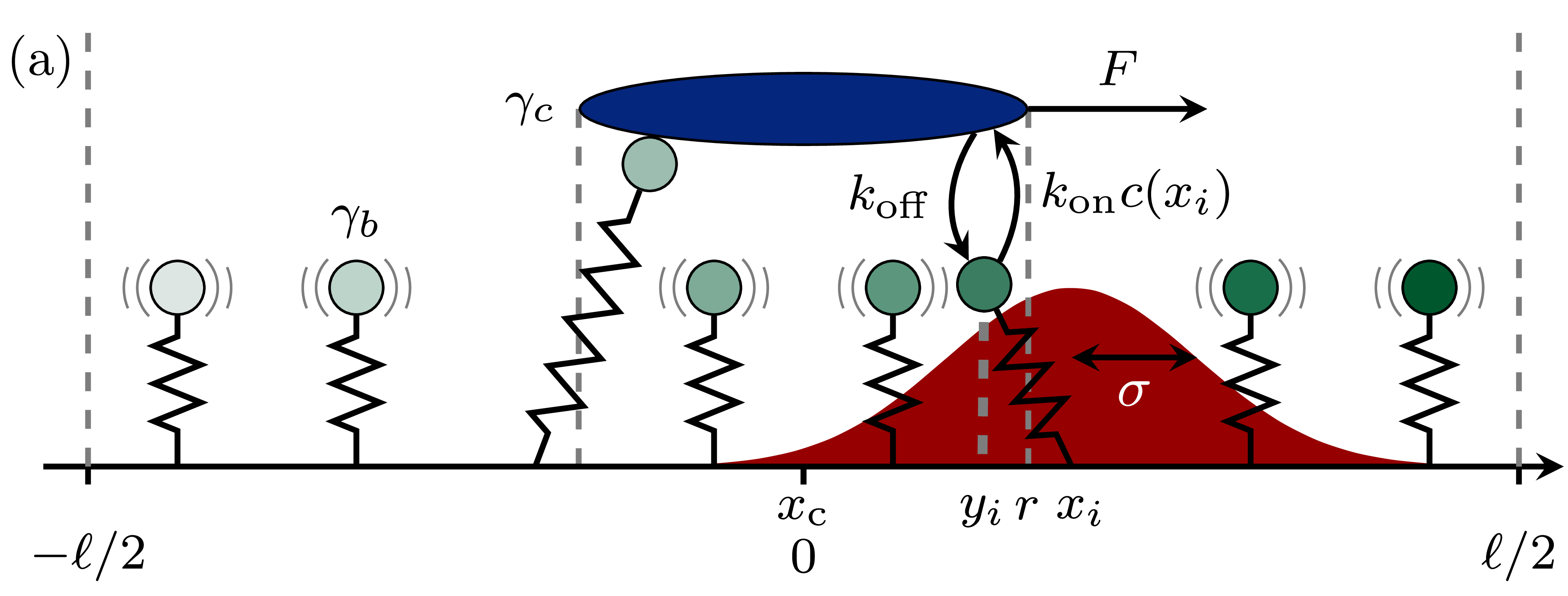}
	\includegraphics[width=\columnwidth]{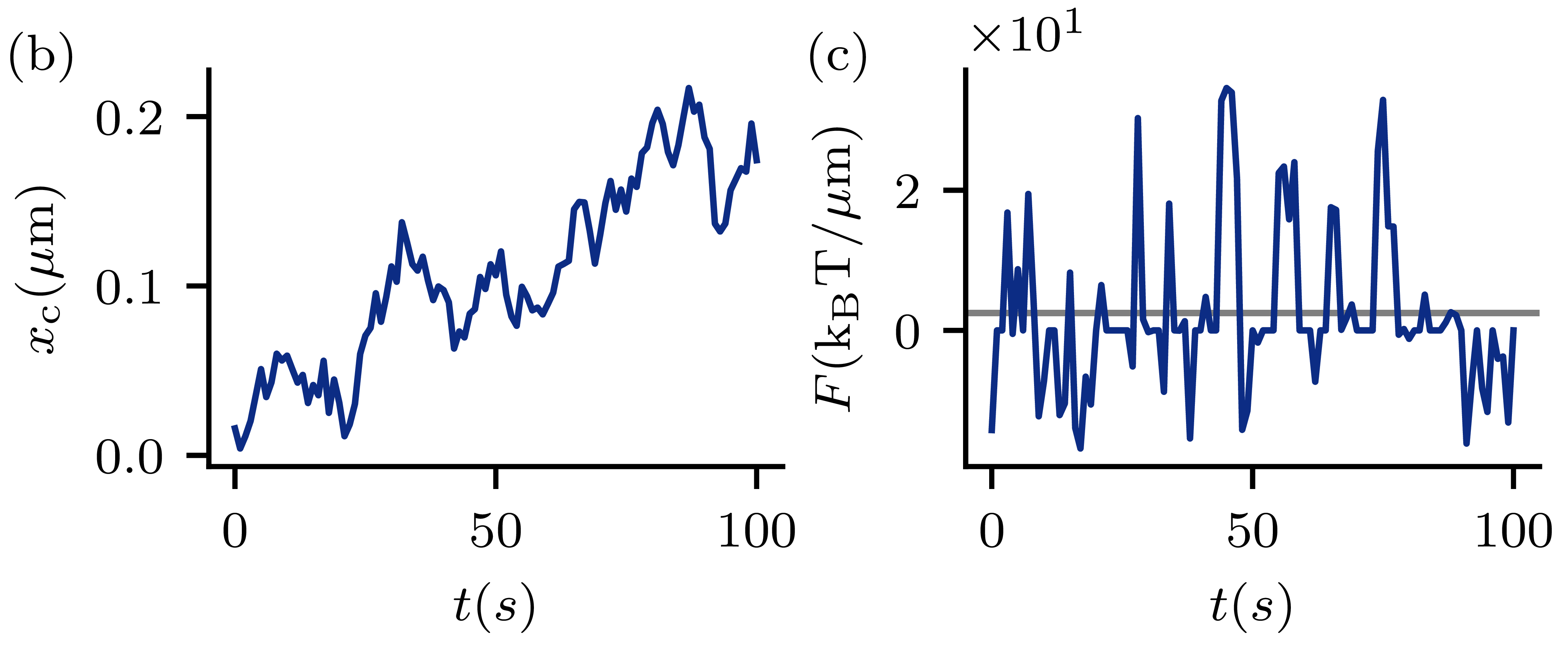}
	\caption{{\bf Minimal model for force generation by DNA-relaying}. 
	(a) The relay force $F$ arises from the interactions of the cargo with ParA ATPases bound to the chromosome, represented by a set of chromosomal elements modelled as a bead-spring system with an associated ParA concentration (indicated by the green tone). 
	We assume that the ParA gradient is co-moving with the cargo. 
	Chromosomal elements fluctuate due to thermal energy, with the magnitude of the fluctuations, $\sigma = \sqrt{k_BT/k}$ (red Gaussian). (b,c) Cargo trajectory (b) and the corresponding DNA-relay force (c) obtained from the numerical solution of Eq.~\eqref{eq:EquationOfMotion} using Brownian dynamics simulation. The horizontal line shows the time-averaged value of $F$.}
	\label{fig:Model}
\end{figure}
To elucidate force generation by the DNA-relay mechanism~\cite{Lim2014,Surovtsev2016a}, we study a minimal model obtained by reducing the full complexity of the partitioning system to key elements important for DNA-relaying (Fig.~\ref{fig:Model}a).
Our one-dimensional model consists of the cargo and ParA-bound chromosomal elements. 
To account for the chromosomal dynamics in a simplified manner, the chromosome is modelled as a set of fluctuating elastic springs.
In ParAB\textit{S}-like partitioning systems, the ATPase ParA detaches from the chromosome at the cargo due to stimulation of ATP hydrolysis by ParB, and can only rebind to the chromosome upon ATP binding and dimerization.
This dynamics results in a ParA gradient propagating with the cargo, as was shown for an \textit{in vitro} reconstituted partitioning system \cite{Hwang2013,Vecchiarelli2014}.
Instead of modeling the ParA dynamics explicitly, we use this observation by imposing a co-moving ParA gradient on the cargo.

Specifically, the cargo is represented as a line segment of length $2r$ with a reaction radius $r$, and chromosomal regions are described in a coarse-grained way as a set of $N_\text{tot}$ beads, equally spaced along a domain of length $\ell$ (Fig.~\ref{fig:Model}a). 
Each bead is tethered to a fixed position by a spring with stiffness $k$, thermally fluctuating with amplitude $\sigma=\sqrt{k_{\text{B}} T/k}$.
The ParA concentration associated with a chromosomal bead at a distance $x_i$ from the cargo is set to $c(x_i)=mx_i+c_0$.
Cargo and chromosomal elements interact: beads within the reaction radius of the cargo bind with rate $k_{\text{on}}c(x)$. 
Cargo-bound beads unbind with rate $k_{\text{off}}$. 
Importantly, due to the elasticity of the DNA, cargo-bound chromosomal elements exert a force on the cargo. 
We describe the resulting cargo motion by an overdamped Langevin equation
\begin{equation}\label{eq:EquationOfMotion}
\gamma_{\rm c} \frac{dx_{\rm c}}{dt}= k\sum_i (x_{i}-y_i) + \sqrt{2\gamma_{\rm c} k_{\rm B} T}\eta(t),
\end{equation}
where $x_{\rm c}$ is the cargo position and the index $i$ runs over all cargo-bound chromosomal elements with rest position $x_i$ and bead position $y_i$.
The white noise term $\eta(t)$ satisfies $\langle \eta(t) \rangle=0$ and $\langle \eta(t)\eta(t') \rangle=\delta(t-t')$, and  $\gamma_{\rm c}$ is the friction coefficient of the cargo in the cytoplasm. 

Our goal is to calculate the steady-state DNA-relay force on the cargo for a co-moving ParA gradient.
To compute the steady-state DNA-relay force using a finite chromosomal domain of size $\ell$, we employ periodic boundary conditions, such that there are always $N_\text{tot}$ chromosomal elements the cargo could interact with~\cite{SupplementalMaterial}. 
For $\sigma \gg \ell$, the limited number of chromosomal elements becomes important, allowing us to study finite system size effects. In contrast, if $\sigma \ll \ell$, this model is effectively identical to one with an infinite system size. 

To facilitate further theoretical analysis we recast variables and system parameters in a non-dimensional form using the system size $\ell$ as a characteristic length, $x\rightarrow x \ell$, and the unbinding time $1/k_{\text{off}}$ as characteristic time scale, $t\rightarrow t/ k_{\text{off}}$. Using this non-dimensionalized form, we identify four key parameters that dictate the system's dynamics:
The binding propensity $c_0 k_{\text{on}}/k_{\text{off}} \rightarrow c_0$ characterizes the on/off kinetics between the cargo and ParA; the concentration gradient $m \ell/c_0 \rightarrow m$ describes the asymmetry of the ParA gradient on the chromosome; $\sigma/ \ell\rightarrow \sigma$ sets the magnitude of chromosomal fluctuation relative to system size; and the cargo friction coefficient $\gamma_{\text{c}} k_\text{off}\ell^2/(k_\text{B}T) \rightarrow \gamma_{\text{c}}$ provides a measure for how susceptible the cargo is to DNA-relay forces~\cite{SupplementalMaterial}.

\begin{figure}[t!]
	\centering
	\includegraphics[width=\columnwidth]{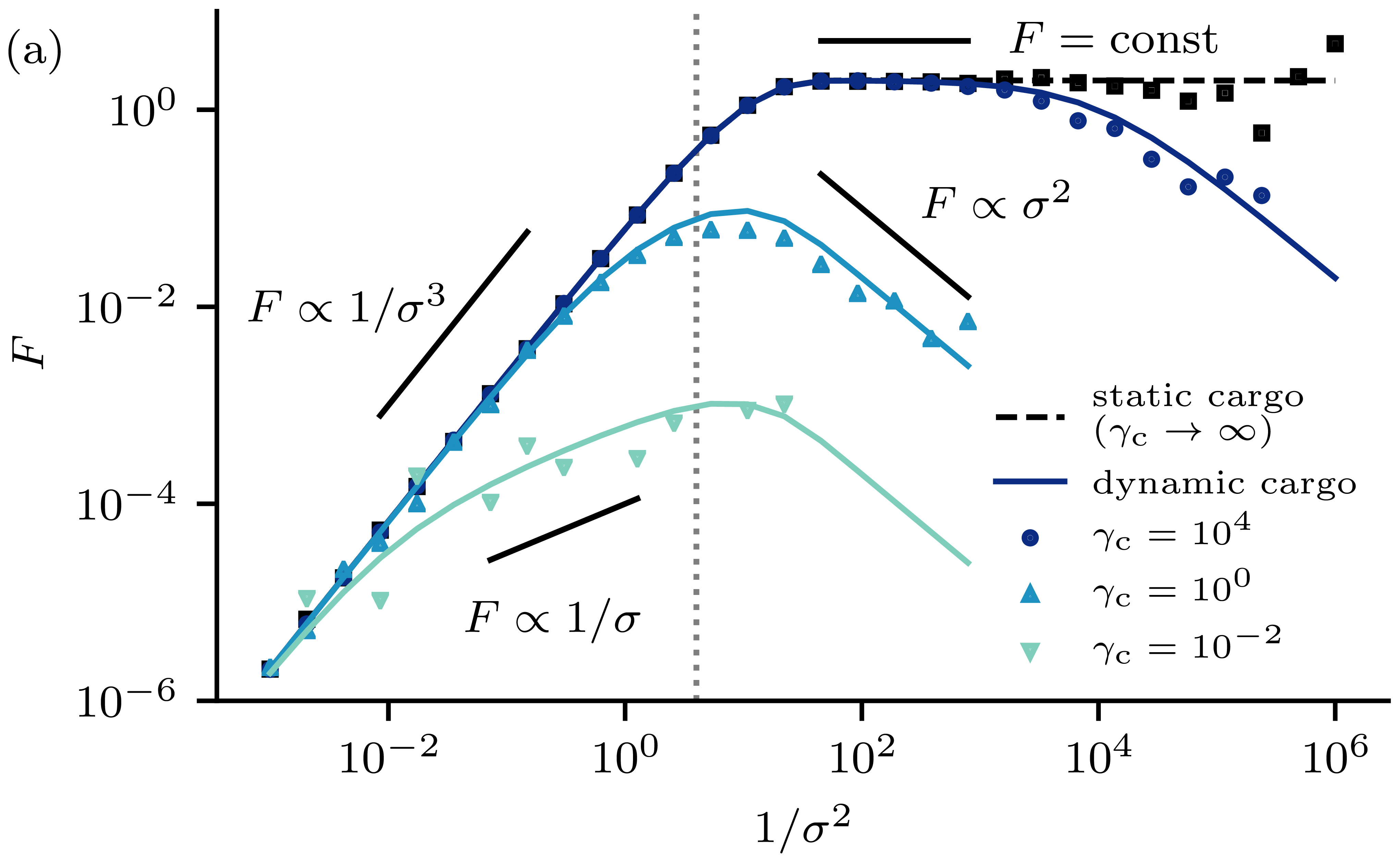}
	\includegraphics[width=\columnwidth]{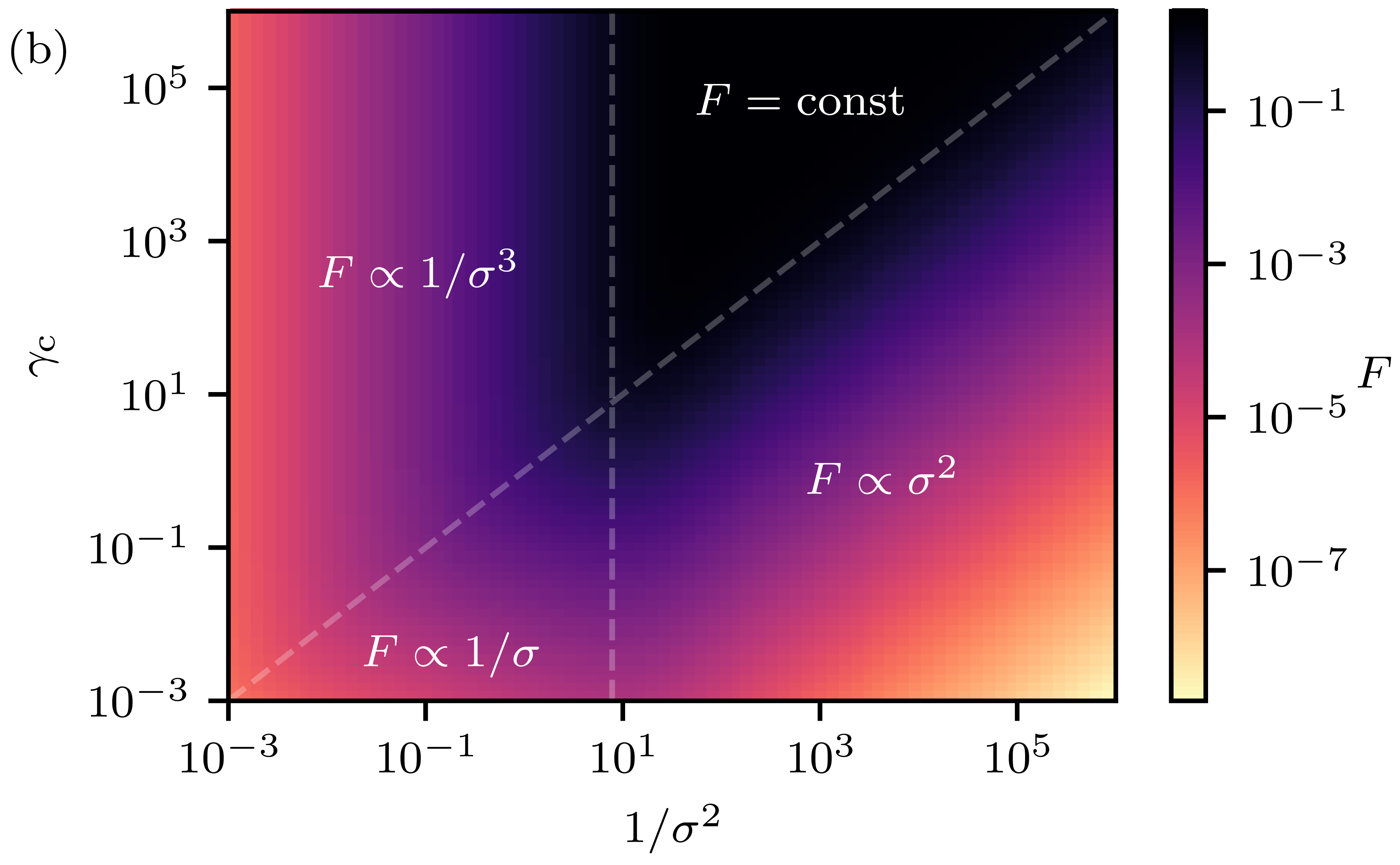}
	\caption{{\bf Average relay force $F$ in the weak-binding limit ($c_0\ll1$) for different values of the friction coefficient $\gamma_{\text{c}}$ of the cargo in the cytoplasm and the magnitude of chromosome fluctuations $\sigma$.} (a) We compare results from simulations (dots) with theory (lines), obtained from Eq.~\eqref{eq:ForceWeakBindingLimit} and Eq.~\eqref{eq:DynamicCargoForceExplicit} for a static (black) and moving cargo (blue). The dotted vertical line at $\sigma=1/2$ separates the different force generation regimes.
	(b) Phase diagram of force generation regimes.}
	\label{fig:DynamicCargoForce}
\end{figure}

Using Brownian dynamics simulations (Fig.~\ref{fig:Model}b,c) we find distinct force-generation regimes depending on the magnitude of chromosomal fluctuations $\sigma$ and the cytoplasmic friction coefficient $\gamma_{\rm c}$ of the cargo, each characterized by a different dependence on $\sigma$ (Fig.~\ref{fig:DynamicCargoForce}). 
While we observe maximal force under stalling conditions ($\gamma_{\text{c}}\rightarrow\infty$), the system's behavior changes drastically for a moving cargo (finite $\gamma_{\text{c}}$). 
Interestingly, in this parameter range we find a maximum in the force at intermediate $\sigma$, suggesting an optimal operating regime for this transport mechanism.

To provide conceptual insight into the DNA-relay mechanism, we develop an analytical theory to calculate the relay force on the cargo. Specifically, we derive an approximation for the relay force
\begin{equation}\label{eq:DNARelayForce}
F=\frac{1}{\sigma^2}\sum_i (x_{i}-y_i),
\end{equation}
which reveals how microscopic system parameters control the DNA-relay mechanism.
To obtain an explicit analytical expression, we consider the average relay force, and use a continuum approximation
\begin{align}
\notag
F &= \frac{1}{\sigma^2}\int_{-1/2}^{1/2} \mathrm{d}x \int_{-r}^{r} \mathrm{d}y \, n(x, y, t) (x-y) \\
&= \int_{-1/2}^{1/2}  \mathrm{d}x \, f(x,t).
\label{eq:Force}
\end{align}
We moved to the cargo frame of reference, introduced the density $n(x,y,t)$ of cargo-bound chromosomal elements with a rest position $x$ and binding position $y$ at the cargo, and defined the force density
\begin{equation}\label{eq:ForceDensity}
f(x,t) = \frac{1}{\sigma^2} \int_{-r}^r \mathrm{d}y \, n(x,y,t) (x-y). 
\end{equation}
Thus, the relay force can be understood by studying the force density $f$, for which we need to calculate $n(x,y,t)$.

The dynamics of the density $n(x,y,t)$ is described by
\begin{multline}\label{eq:MasterEquation}
\partial_t n(x,y,t) - v(n,t) \partial_x n(x,y,t) = \\
c(x) \phi(y;x,\sigma) (N_\text{tot}-n(x,t)) - n(x,y,t).
\end{multline}
For a static cargo ($v=0$), the temporal change in  $n$ is determined only by a gain and a loss term, corresponding to binding to and unbinding from the cargo. 
For a binding event, a chromosomal bead needs to move within the reaction radius of the cargo. 
We describe the position $y$ of an unbound bead as a Gaussian random variable with mean $x$ and variance $\sigma^2$. The probability that a bead with rest position $x$ is at position $y \in [-r,r]$ is thus given by the Gaussian probability density function $\phi(y;x,\sigma)$ (Fig.~\ref{fig:Model}a).
This is justified under weak chromosome-cargo interactions, i.e.~whenever the decorrelation time $\tau_{\text{corr}}=\sigma^2 \gamma_{\text{b}}N_\text{tot}$ is much smaller than the binding time $\tau_{\text{bind}}=1/c_0$.
A binding event takes place stochastically with a rate $c(x)(N_\text{tot}-n(x,t))$, accounting for the finite density of chromosomal elements available for binding, where $c(x) = c_0 (1+mx)$ denotes the dimensionless ParA concentration. The total density of cargo-bound chromosomal beads with rest position $x$ can be obtained by integrating the density $n(x,y,t)$ over all possible binding positions $y$ on the cargo:
\begin{equation}
n(x,t) = \int_{-r}^r n(x,y,t) \, \mathrm{d}y.
\end{equation}
Unbinding is described by a constant detachment rate, set by the last term in Eq.~\eqref{eq:MasterEquation}.
Finally, when $v\neq0$ the temporal evolution of $n$ also includes an advection term to account for cargo motion.

We expect the weak-binding limit ($c_0 \ll 1$) to be the biologically relevant parameter regime in this model, because of the high ParA turnover rate caused by ParB-induced ATP hydrolysis of ParA and subsequent detachment of ParA from the cargo \cite{Vecchiarelli2010}. 
Henceforth, we thus consider only this limit, for which saturation effects of the cargo by bound chromosomal elements are negligible.
For completeness, we provide our results for the strong-binding limit~\cite{SupplementalMaterial} and find that the conceptual insights gained from the weak-binding limit largely apply.

Having established a theoretical framework to study force generation by DNA-relaying, we first consider the case of a static cargo ($v=0$). Put simply, we compute the stalling force of the cargo. This static case allows us to study basic features of the force generation mechanism and provides insights that will also be relevant for the moving cargo scenario. 
We first calculate the steady-state solution of Eq.~\eqref{eq:MasterEquation}, and with this an expression for the steady-state force density~\cite{SupplementalMaterial}:
\begin{equation}
\label{eq:ForceDensityWbl}
f(x) = c(x) (\phi(x;r,\sigma)-\phi(x;-r,\sigma)).
\end{equation}
This expression for the force density constitutes one of our key findings and allows us to understand 
how the DNA-relay force is generated and how it depends on system parameters.

\begin{figure}[t!]
	\centering
	\includegraphics[width=\columnwidth]{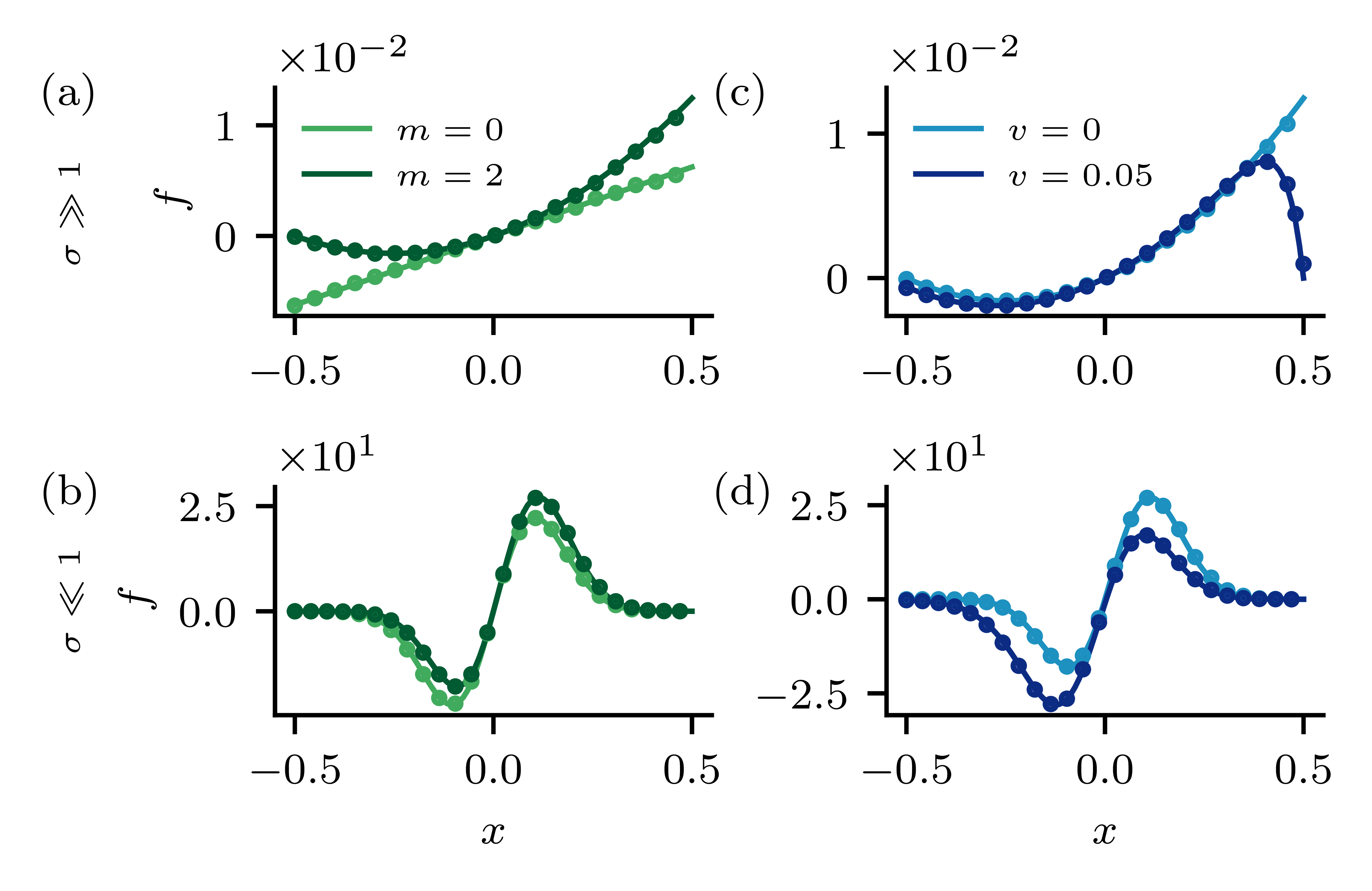}
	\caption{ {\bf The influence of the ParA concentration gradient $m$ and the cargo velocity $v$ on the force density $f(x)$}. (a,b) $f(x)$ for a static cargo given by Eq.~\eqref{eq:ForceDensityWbl} without ($m=0$) and with ($m=2$) a ParA gradient. (c,d) $f(x)$ for a static ($v=0$) and a moving ($v=0.05$) cargo both with $m=2$. The force density for a moving cargo is obtained numerically. We compare results from simulations (dots) and our theoretical results (lines). Note that the dark green and the light blue curves in (a,c) and (b,d) show the same data.}
	\label{fig:Distributions}
\end{figure}

The force density encodes the contribution of a chromosomal element with rest position $x$ to force generation.
Intuitively, this force density is determined by the interplay between how likely it is for a chromosomal element to bind to the cargo and how much force is exerted on the cargo in this configuration.
In the limit $\sigma \gg 1$, chromosomal beads exhibit strong fluctuations, and without a ParA gradient ($m = 0$) every bead thus has approximately the same binding probability. 
Here, only the distance of a chromosomal element from the cargo matters for force generation and therefore the force density increases linearly with the distance of the bead from the cargo (Fig.~\ref{fig:Distributions}a, light green).
Because of the symmetry of $f(x)$, forces exerted on the cargo from chromosomal elements positioned behind and in front of the cargo cancel, such that no net force is generated. By contrast, if the ParA concentration on the beads increases towards the right $(m > 0)$, beads in front of the cargo are more likely to bind to the cargo than those behind. 
Hence, the force density profile becomes asymmetric, resulting in a net positive force (Fig.~\ref{fig:Distributions}a, dark green).
In the regime $\sigma \ll 1$ there is a non-uniform probability for chromosomal beads to bind to the cargo. 
While chromosomal elements far from the cargo are less likely to bind, they generate the largest force contribution. Consequently, the force density peaks at an intermediate position between the cargo edge and the system boundary (Fig.~\ref{fig:Distributions}b).
Again, in the presence of a ParA gradient $f(x)$ becomes asymmetric, resulting in a net force on the cargo. 
In all cases, our analytical predictions for the force density are in accord with Brownian dynamics simulations.

Having analyzed the steady-state force density $f(x)$, we next evaluate the cargo stalling force $F_{\text{sc}}$ in the weak-binding limit using Eq.~\eqref{eq:Force}:
\begin{equation}
\label{eq:ForceWeakBindingLimit}
F_{\text{sc}} = m c_0 \int_{-1/2}^{1/2} \mathrm{d}x \,x (\phi(x;r,\sigma) - \phi(x;-r,\sigma)) 
\end{equation}
Upon performing this integral, we obtain the dependence of the cargo stalling force on $\sigma$ (Fig.~\ref{fig:DynamicCargoForce}).
Remarkably, for $\sigma \ll1$ we find that $F_{\text{sc}}$ is independent of $\sigma$. Upon increasing $\sigma$, more chromosomal elements are recruited to contribute to force generation. 
However, this increase in participation is precisely compensated by the softening of the springs resulting in a stiffness independent DNA-relay force $F_{\text{sc}}={\rm const}$.
For $\sigma \gg 1$, we obtain $F_{\text{sc}}\propto1/\sigma^{3}$. 
Here, the finite size of the system affects force generation. Due to the limited number of beads, the softening of the springs can not be compensated anymore by an increased amount of beads interacting with the cargo. Therefore, the force on the cargo decreases.

To understand force generation for a dynamic cargo, we first consider the case of a cargo that moves with an imposed velocity $v$. To this end, we study the steady-state force density, which determines the relay force $F(v)$.
We calculate the steady-state solution of Eq.~\eqref{eq:MasterEquation} for a fixed velocity $v$ and obtain the corresponding force density $f(x)$ using Eq.~\eqref{eq:ForceDensity}.
We observe that, for $v>0$, weight of the binding profile is relocated from the leading (right) to the lagging (left) side of the cargo (Fig.~\ref{fig:Distributions}c,d, dark blue). 
This can be understood intuitively: In the case of a dynamic cargo, the forward movement of the cargo and the finite time a chromosomal bead is attached to the cargo (on average $1/k_\text{off}$), result in an increased amount of chromosomal beads pulling the cargo backwards. 

Interestingly, we find that a moving cargo experiences the force
\begin{equation}
	F(v)=F_{\text{sc}} -v\frac{1}{\sigma^2} N_{\text{sc}},
\end{equation}
which has two contributions: the static relay force and an additional force term linear in $v$. 
This term can be interpreted as an emergent friction force with the friction coefficient $\gamma_{\text{e}}=\frac{1}{\sigma^2} N_{\text{sc}}=\frac{1}{\sigma^2} 2 r c_0$, where $N_{\text{sc}}$ denotes average number of cargo bound beads for a static cargo~\cite{SupplementalMaterial}.

Next, we use this result for imposed motion to obtain the DNA-relay force exerted on a cargo that moves autonomously due to diffusion and the interactions with ParA-bound beads.
First, we self-consistently determine the velocity $v$ of a self-propelled cargo using force balance
$\gamma_{\text{c}} v = F(v)$.
From this analysis, we obtain an explicit expression for the generated force associated to this translocation velocity
\begin{equation}\label{eq:DynamicCargoForceExplicit}
F = \frac{F_{\text{sc}}}{1+\frac{\gamma_{\text{e}}}{\gamma_{\text{c}}}}. 
\end{equation}
Interestingly, the force on an autonomously moving cargo can be entirely calculated from quantities obtained for a static cargo.

The interplay of self-propulsion and emergent friction force gives rise to four distinct force generation regimes, as depicted in the phase diagram in Fig.~\ref{fig:DynamicCargoForce}b. As in the static limit, we can distinguish force generation for small and large chromosomal fluctuations. Importantly however, the qualitative dependencies on the strength of the chromosome fluctuations can differ because of the emergent friction force. In the limit where the cytoplasmic friction dominates the emergent friction, $\gamma_{\text{c}}\gg\gamma_{\text{e}}$, the dynamic relay force is well approximated by the static relay force (Fig.~\ref{fig:DynamicCargoForce}a, black line).
Upon lowering the cytoplasmic friction slightly, the emergent friction only reduces force generation for small $\sigma$. 
Here, the $\sigma$-dependence of the emergent friction, $\gamma_{\text{e}}\propto1/\sigma^2$, combines with the constant static cargo force to $F\propto\sigma^2$ (Fig.~\ref{fig:DynamicCargoForce}a, dark blue line).
Upon lowering $\gamma_{\text{c}}$ further the emergent friction also influences the regime $\sigma\gg1$. For this parameter regime, the decrease in driving and friction force with increasing $ \sigma$ combine to $F\propto1/\sigma$ (Fig.~\ref{fig:DynamicCargoForce}a, light blue line). In the limit $\sigma \rightarrow \infty$, we find that the relay force vanishes, as for a static cargo. In all cases, we find that our analytical predictions agree well with the Brownian dynamics simulations.

Our work complements previous studies on numerically and phenomenologically modeling cargo motion in ParAB\textit{S}-like systems~\cite{Lim2014, Jindal2019,Hu2015,Hu2017,Hu2017a,Surovtsev2016a, Bergeler2018} by providing an analytical microscopic theory for force generation by DNA-relaying. 
It is still debated whether the main contribution to force generation in
ParAB\textit{S} systems derives from chromosome elasticity (DNA-relay force)~\cite{Wiggins2010, Lim2014,Hu2015,Hu2017,Hu2017a} or chemophoresis~\cite{Jindal2015,Sugawara2011, Walter2017}.
We contribute to this open question by developing a quantitative mechanistic theory. Our analytical predictions for the dependence of the DNA-relay force on microscopic parameters could be tested in \textit{in vitro} experiments with a stiffness controlled DNA-carpet~\cite{Vecchiarelli2014}.
In future work, our framework can serve as a starting point for further investigations of force generation in ParAB\textit{S} systems with complex ParA dynamical patterns~\cite{Ringgaard2009} and non-equilibrium activity in the chromosome~\cite{Gnesotto2018,Weber2012}. Our theory might also be useful more generally for macroscopic cargo translocation driven by stochastic interactions \cite{Sabass2010,Srinivasan2009}.

\begin{acknowledgments}
	This research was funded by the Deutsche Forschungsgemeinschaft (DFG, German
	Research Foundation, Project 269423233 - TRR 174). C.H. thanks the Max Planck Institute for the Physics of Complex Systems, Dresden (Germany) for hospitality.
\end{acknowledgments}

\bibliography{./references.bib}
\end{document}


\title{Supplemental Information for ``Theory of Active Intracellular Transport by DNA-relaying''}

\author{Christian Hanauer}
\thanks{These authors contributed equally to this work.}
\affiliation{Arnold-Sommerfeld-Center for Theoretical Physics and Center for
	NanoScience, Ludwig-Maximilians-Universit\"at M\"unchen,
	D-80333 M\"unchen, Germany.}

\author{Silke Bergeler}
\thanks{These authors contributed equally to this work.}
\affiliation{Arnold-Sommerfeld-Center for Theoretical Physics and Center for
	NanoScience, Ludwig-Maximilians-Universit\"at M\"unchen,
	D-80333 M\"unchen, Germany.}

\author{Erwin Frey}
\affiliation{Arnold-Sommerfeld-Center for Theoretical Physics and Center for
	NanoScience, Ludwig-Maximilians-Universit\"at M\"unchen,
	D-80333 M\"unchen, Germany.}

\author{Chase P. Broedersz}
\email{c.p.broedersz@vu.nl}
\affiliation{Arnold-Sommerfeld-Center for Theoretical Physics and Center for
	NanoScience, Ludwig-Maximilians-Universit\"at M\"unchen,
	D-80333 M\"unchen, Germany.}
\affiliation{Department of Physics and Astronomy, Vrije Universiteit Amsterdam, 1081 HV Amsterdam, The Netherlands}  
 
\pacs{}
\date{\today}

\maketitle
\noindent
	\section{Brownian dynamics simulation}\label{sec:section1}
	\begin{figure}[h!]
		\centering
		\includegraphics[width=90mm]{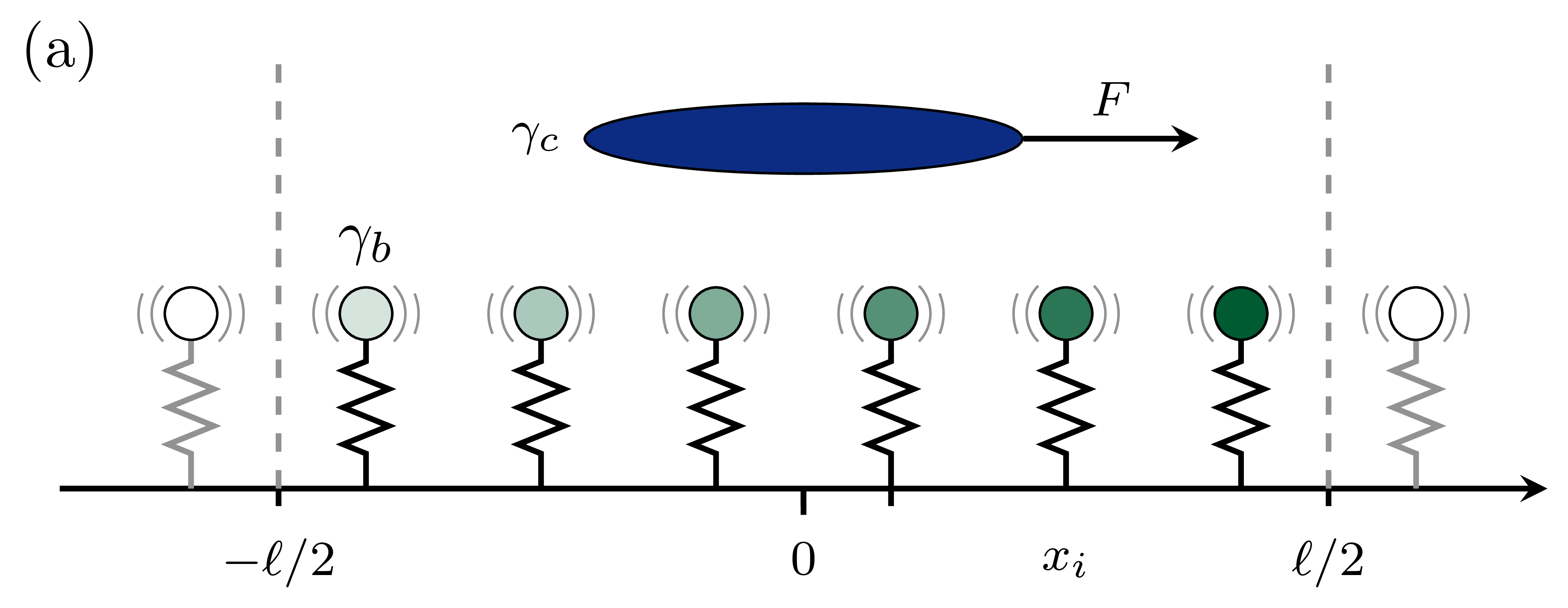}
		\includegraphics[width=58mm]{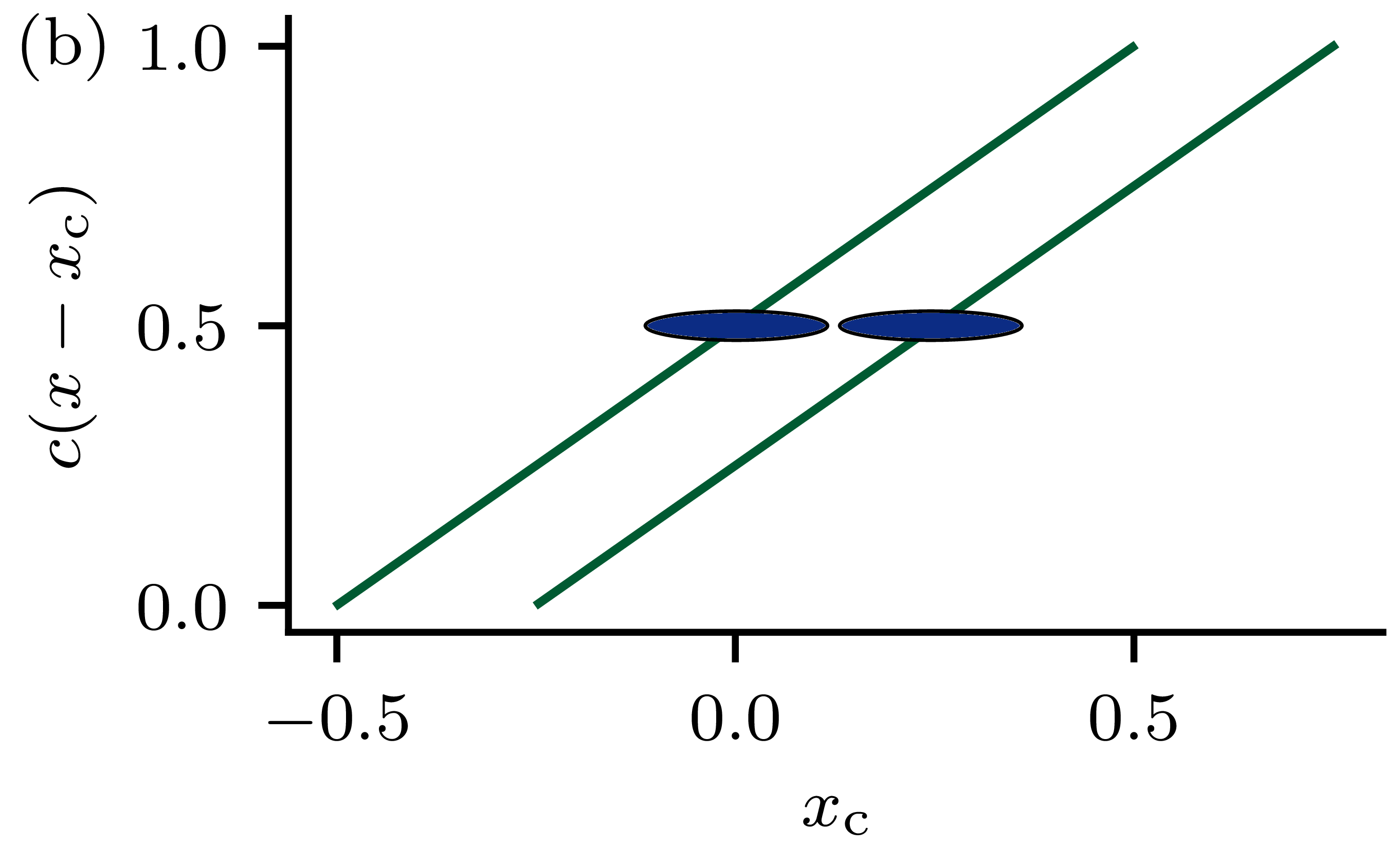}
		\caption{{\bf Schematics to illustrate the implementation of our model to investigate the steady-state DNA-relay force exerted on the cargo.} (a) To obtain steady-state cargo forces without simulating large chromosomal regions, we consider a finite number of beads $N_\text{tot}$ around the cargo, which correspond to a chromosomal region of size $\ell$. When the cargo moves to the right, beads further away from the cargo move outside of the region $[x_c-\ell/2,x_c+\ell/2]$. These beads are reintroduced in front of the cargo such that the cargo is always surrounded by the same amount of beads. The ParA concentration on the chromosomal elements is indicated by the green tone. (b) We impose a fixed, linear ParA concentration gradient on the beads in the co-moving reference frame of the cargo.}
		\label{fig:BoundaryConditions}
	\end{figure}
	\noindent We perform one-dimensional Brownian dynamics simulations of cargo motion. The cargo position, $x_{\text{c}}$ is described by the overdamped Langevin equation 
	\begin{equation}\label{eq:LangevinEquation}
	\gamma_{\text{c}} \frac{dx_{\text{c}}}{dt}= F + \sqrt{2\gamma_{\text{c}} k_{\text{B}} T}\eta(t),
	\end{equation}
	where $\gamma_{\text{c}}$ refers to the cytosolic friction coefficient of the cargo and the white noise term $\eta(t)$ satisfies $\langle \eta(t) \rangle=0$ and $\langle \eta(t)\eta(t') \rangle=\delta(t-t')$. The DNA-relay force term reads
	\begin{align}
	F = -k \sum_i &(y_{i}-x_{i}),
	\end{align}
	where the sum runs over all cargo-bound chromosomal elements, $x_i$ refers to the rest position and $y_i$ to the position of bead $i$. We use the Euler-Maruyama scheme \cite{Kloeden2011} to find a numerical solution to Eq. \eqref{eq:LangevinEquation} at time steps $t_k=k \Delta t$ with $k=0,\dots,N_{\text{steps}}$ via the relation
	\begin{equation}\label{eq:EMScheme}
	x_{k+1} = x_k + \frac{F(x_k)}{\gamma_{\rm c}}\Delta t + \sqrt{\frac{2 k_{\text{B}}T}{\gamma_{\rm c}}}\, \Delta W_k,
	\end{equation}
	where the independent Wiener increments $\Delta W_k$ are Gaussian random variables with $\Delta W_k=\mathcal{N}(0,\Delta t)$. To reduce computational complexity we do not explicitly use a Brownian dynamics scheme to simulate the positions of the beads. Instead, we draw the current bead position $y_i$ from the Gaussian distribution $\mathcal{N}(x_i,\sigma)$.
	
	Given that a bead is in the cargo reaction radius ($|x_c-y_i|\leq r$), we simulate the reactions between cargo and bead using a simple stochastic algorithm. First, we generate a uniform random number $\nu$ in the interval $(0,1)$. If $\nu<k_{\text{on}}c(x_{i})\Delta t$, the bead attaches to the cargo. 
	If $\nu\geq k_{\text{on}}c(x_i)\Delta t$, no reaction takes place in this time step.
	Once a bead is cargo-bound, an unbinding event can occur at any time step and is simulated with the same algorithm explained above, but with the rate $k_\text{off}$. We choose $\Delta t$ sufficiently small, such that it is very unlikely that more than one stochastic event takes place during $\Delta t$.
	
	Our goal is to obtain the steady-state DNA-relay force and its dependence on the model parameters without considering boundary effects of the chromosome.	Since the simulation of an infinitely large system is not possible, we approximate an infinitely large system by using a finite chromosomal region around the cargo. When the cargo moves to the right, beads to the left of the cargo can move outside of the region $[x_c-\ell/2 ,x_c + \ell/2]$ around the cargo (Fig.~\ref{fig:BoundaryConditions}). These beads are then reintroduced in front of the cargo, such that the cargo is always surrounded by the same $N_\text{tot}$ beads. 
	Furthermore, we assume that the ParA concentration gradient on the beads is co-moving with the cargo position, such that the cargo experiences the same gradient when it is moving through the system:
	\begin{equation}
	c(x) = m(x-x_{\text{c}}) + c_0
	\end{equation}	
	\begin{table}[h]
		\centering
		\caption[Summary of model parameters]{Summary of model parameters. If not otherwise stated, we use the parameter values listed here for our simulation. The notation $[\cdot,\cdot]$ indicates a range of parameters.}
		\label{tab:ModelParameters}
		\begin{ruledtabular}
			\begin{tabular}{l l l l l l l l l l l}
				Parameter & Symbol & Unit & Fig.~1 & Fig.~2a & Fig.~2b & Fig.~3 & Fig.~S2 & Fig.~S3 \\
				Cargo friction coefficient & $\gamma_{\text{c}}$ & $\rm k_{\rm B} T$ \si{\second\per\micro\squared\meter} 
				& $10^4$ & $10^{-2},1,10^4$ & $[10^{-3},10^{6}]$& $10^4$&  $10^4$ & $1$\\
				Spring constant & $k$& $\rm k_{\rm B}T$\si{\per\micro\squared\metre}
				& 100 & $[10^{-3},10^{6}]$ & $[10^{-3},10^{6}]$ & $0.1,100$ & $0.1,10^4$ & $[10^{-3},10^6]$ \\
				Binding rate & $k_{\text{on}}$& \si{\micro\meter\per\second}
				& $0.01$ & $0.01$ & $0.01$ & $0.01$& $[10^{3},10^4,10^5]$  & $[10^{-4},10^3,10^4]$\\
				Unbinding rate & $k_{\text{off}}$& \si{\per\second} & $1$ & $1$ & $1$ & $1$& $1$  & $0.01$ \\
				Offset concentration & $c_{\text{0}}$& \si{\per\micro\meter}
				& $1$ & $1$ & $1$ & $1$& $1$ & $1$\\
				Concentration gradient & $m$ & \si{\per\micro\meter\squared} 
				&$2$ & $2$ & $2$ & $[0,2]$& $2$ & $2$\\
				Number of beads & $N_{\text{tot}}$& 1 & $1000$ & $1000$ & $1000$ & $1000$  & $1000$ & $1000$\\ 
				Cargo radius & $r$& \si{\micro\metre} & $0.05$ & $0.05$ & $0.05$ & $0.05$  & $0.05$ & $0.05$\\
				System length & $\ell$& \si{\micro\metre} & $1$ & $1$ & $1$ & $1$& $1$ & $1$ \\
			\end{tabular}
		\end{ruledtabular}
	\end{table}
	
	\section{Non-dimensionalization}
	\noindent Here we describe how we choose the time and length scales as well as the characteristic scale of the density $n(x,y,t)$ to arrive at a non-dimensional differential equation for $n(x,y,t)$ (Eq.~(5) in the main text). In dimensional units, the time evolution of the density $n$ is given by
	\begin{equation}
	\partial_t n(x,y,t) - v(n,t) \partial_x n(x,y,t) = k_\text{on} c(x) \phi(y;x,\sigma) \left(\frac{N_\text{tot}}{\ell} - n(x,t)\right) - k_\text{off} n(x,y,t),
	\end{equation}
	together with
	\begin{equation}
	\gamma_{\text{c}} v = F\qquad\text{and}\qquad F = k\int_{-\ell/2}^{\ell/2} \mathrm{d}x \int_{-r}^{r} \mathrm{d}y \, n(x, y, t) (x-y).
	\end{equation}
	The ParA concentration gradient is denoted by $c(x) = mx + c_0$ and the position $x$ of a chromosomal element with rest position $y$ is described by the Gaussian probability density $\phi(y;x,\sigma)$.
	
	Next, we express lengths in terms of the system size $\ell$, $x = \tilde{x} \ell$, and time in terms of the average time until a cargo-bound bead detaches from the cargo, $t = \tilde{t}/k_\text{off}$:
	\begin{equation}
	k_\text{off} \partial_{\tilde{t}} n(x,y,t) - \frac{v(n,\tilde{t})}{\ell} \partial_{\tilde{x}} n(x,y,t) = k_\text{on} (m \ell \tilde{x} + c_0) \frac{\phi(\tilde{y};\tilde{x},\tilde{\sigma})}{\ell} \left(\frac{N_\text{tot}}{\ell} - n(x,t)\right) - k_\text{off} n(x,y,t).
	\end{equation}
	The density $n(x,y,t)$ becomes dimensionless by using $n(x,y,t) =  \tilde{n}(\tilde{x},\tilde{y},\tilde{t})/\ell^2$ and therefore $n(x,t) = \tilde{n}(\tilde{x},\tilde{t})/\ell$.
	Multiplying the above equation by $\ell^2/k_\text{off}$ yields 
	\begin{equation}
	\partial_{\tilde{t}} \tilde{n}(\tilde{x},\tilde{y},\tilde{t}) - \tilde{v}(\tilde{n},\tilde{t}) \partial_{\tilde{x}} \tilde{n}(\tilde{x},\tilde{y},\tilde{t}) = \frac{k_\text{on}}{k_\text{off}} (m \ell \tilde{x}  + c_0) \phi(\tilde{y};\tilde{x},\tilde{\sigma}) \left(N_{\text{tot}}- \tilde{n}(\tilde{x},\tilde{t})\right) - \tilde{n}(\tilde{x},\tilde{y},\tilde{t}).
	\end{equation}
	Upon defining $c_0 =\tilde{c}_0 k_\text{off}/k_\text{on}$ and $m=\tilde{m} c_0/\ell$, we obtain:
	\begin{equation}
	\partial_{\tilde{t}} \tilde{n}(\tilde{x},\tilde{y},\tilde{t}) - \tilde{v}(\tilde{n},\tilde{t}) \partial_{\tilde{x}} \tilde{n}(\tilde{x},\tilde{y},\tilde{t}) = \tilde{c}(\tilde{x}) \phi(\tilde{y};\tilde{x},\tilde{\sigma}) \left(N_{\text{tot}}- \tilde{n}(\tilde{x},\tilde{t})\right) - \tilde{n}(\tilde{x},\tilde{y},\tilde{t}),
	\end{equation}
	with $\tilde{c}(\tilde{x}) = \tilde{c}_0(1+\tilde{m} \tilde{x})$.
	
	The expression for the relay force, in terms of non-dimensional quantities, reads
	\begin{align}
	F &= \frac{k_{\text{B}}T}{\ell} \frac{1}{\tilde{\sigma}^2}\int_{-1/2}^{1/2} \mathrm{d}\tilde{x} \int_{-\tilde{r}}^{\tilde{r}} \mathrm{d}\tilde{y} \, \tilde{n}(\tilde{x}, \tilde{y}, \tilde{t}) (\tilde{x}-\tilde{y}) \\
	&= \frac{k_BT}{\ell}\tilde{F}.
	\end{align}
	Finally, we find by
	\begin{align}
	\gamma_{\text{c}} v &= F \\
	\frac{k_\text{off} \ell^2}{k_B T}\gamma_{\text{c}} \tilde{v}	&= \tilde{F}
	\end{align}
	that the characteristic scale of the friction coefficient is $k_{\text{B}}T/k_{\text{off}}\ell^2$. 
	In the following, as well as in the main text, we omit the tilde to simplify the notation.	
	\section{Theoretical approximation for DNA-relay force}
	\noindent The time evolution of the density of chromosomal elements $n(x,y,t)$ that are bound to the cargo with rest position $x$ and bead position $y$ is given by:
	\begin{equation}
	\label{eq:nxyt}
	\partial_t n(x,y,t) - v(n,t) \partial_x n(x,y,t) = c(x) \phi(y;x,\sigma) (N_{\text{tot}}-n(x,t)) - n(x,y,t),
	\end{equation}
	with the cargo velocity $v$, the rescaled expression for the ParA density, $c$, and the Gaussian probability density function $\phi$.
	The density $n(x,t)$ is the density of all cargo-bound beads with rest position $x$: 
	\begin{equation}
	n(x,t) = \int_{-r}^r n(x,y,t) \, \mathrm{d}y.
	\end{equation}
	Integrating Eq.~\eqref{eq:nxyt} over all possible bead-binding positions at the cargo, yields an equation for the density of bound beads with rest position $x$:
	\begin{align}
	\nonumber
	\partial_t n(x,t) - v(n,t) \partial_x n(x,t) &= c(x) \left(\int_{-r}^r \phi(y;x,\sigma) \, \mathrm{d} y\right) (N_{\text{tot}}-n(x,t)) - n(x,t) \\
	\label{eq:nxt}
	&= c(x) p_f(x) (N_{\text{tot}}-n(x,t)) - n(x,t),
	\end{align}
	with the finding probability $p_f(x)$. 
	For a rest position $x$, we calculate the integral over all deflections weighted by $n(x,y,t)$:
	\begin{align}
	f(x,t) = \frac{1}{\sigma^2}\int_{-r}^r n(x,y,t) (x-y)\, \mathrm{d}y.
	\end{align}
	To get the total relay force, we need to integrate over all deflections of cargo-bound beads:
	\begin{align}\label{eq:Force}
	F = \int_{-1/2}^{1/2} f(x,t)\, \mathrm{d}x.
	\end{align}
	With Eq.~\eqref{eq:nxyt}, the temporal evolution of the relay force density is given by:
	\begin{align}
	\label{eq:fxt}
	\partial_t f(x,t) =  v(n,t)\left(\partial_x f(x,t) - \frac{1}{\sigma^2}n(x,t)\right) + c(x) (N_{\text{tot}}-n(x,t))\frac{1}{\sigma^2}\int_{-r}^r \phi(y;x,\sigma)(x-y)\,\mathrm{d}y - f(x,t).
	\end{align}
	Eq.~\eqref{eq:nxt} and Eq.~\eqref{eq:fxt} constitute a system of partial differential equations. Next, we obtain steady-state solutions for $n$ and $f$ and use them to find expressions for the amount of cargo-bound beads and the force on a cargo.
	\subsection{Stationary cargo case}
	\noindent For a stationary cargo, i.e. $v = 0$, the steady-state distribution $n(x)$ is given by
	\begin{equation}\label{eq:Density}
	n(x) = \frac{N_{\text{tot}} c(x) p_f(x)}{1+c(x) p_f(x)}.
	\end{equation}
	With this expression, we calculate the steady-state expression for $f$ in the stationary cargo case:
	\begin{align}
	f(x) &= \frac{1}{\sigma^2} c(x) (N_{\text{tot}}-n(x)) \int_{-r}^r \phi(y;x,\sigma)(x-y) \, \mathrm{d}y\\
	&=  \frac{1}{\sigma^2}\frac{N_{\text{tot}}c(x)}{1+c(x)p_f(x)} \int_{-r}^r \phi(y;x,\sigma)(x-y) \, \mathrm{d}y.\label{eq:ForceDensity}
	\end{align}
	Using these results, we compute the average number of bound beads for a stationary cargo
	\begin{equation}
	N_{\text{sc}} = \int_{-1/2}^{1/2} n(x) \, \mathrm{d}x = \int_{-1/2}^{1/2} \frac{N_{\text{tot}}c(x) p_f(x)}{1+c(x)p_f(x)} \, \mathrm{d}x
	\end{equation}
	and the stalling force:
	\begin{equation}
	\label{eq:FscSI}
	F_{\text{sc}} = \int_{-1/2}^{1/2} f(x) \, \mathrm{d}x 
	= \frac{1}{\sigma^2}\int_{-1/2}^{1/2} \mathrm{d}x \, \frac{N_{\text{tot}}c(x)}{1+c(x)p_f(x)} \int_{-r}^r \mathrm{d}y \, \phi(y;x,\sigma)(x-y)
	\end{equation}
	\subsection{Dynamic cargo case}
	\noindent To obtain an expression for the force density in the dynamic cargo case, we first consider a cargo with an externally imposed velocity $v$. 
	Hence, we need to solve the following differential equations to determine the steady-state expressions of the density of cargo-bound beads $n$ and the relay force density $f$:
	\begin{align}
	\label{eq:nxv}
	- v \partial_x n(x,v) &= c(x) p_f(x) (N_{\text{tot}}-n(x,v)) - n(x,v),\\
	\label{eq:Dxxv}
	- v \left( \partial_x f(x,v) - \frac{1}{\sigma^2}n(x,v)\right) &= \frac{1}{\sigma^2}c(x) (N_{\text{tot}}-n(x,v))\int_{-r}^r \phi(y;x,\sigma)(x-y)\,\mathrm{d}y - f(x,v).
	\end{align}
	A numerical solution of these equations is shown in Fig.~\ref{fig:Distributions_sbl}. To determine approximate analytical solutions, we write $n(x,v)$ and $f(x,v)$ as Taylor expansions in the velocity $v$:
	\begin{align}\label{eq:SeriesAnsatz}
	n(x,v) &= n_0(x) + v\,n_1(x) + \mathcal{O}(v^2) , \\
	f(x,v) &= f_0(x) + v\, f_1(x) + \mathcal{O}(v^2).
	\end{align}
	We insert these expressions into Eq.~\eqref{eq:nxv} and \eqref{eq:Dxxv}. 
	To zeroth order in $v$ we get the following two equations
	\begin{align}
	0 &= c(x) p_f(x) (N_{\text{tot}}-n_0(x))-n_0(x),\\
	0 &= \frac{1}{\sigma^2}c(x) (N_{\text{tot}}-n_0(x))\int_{-r}^r \phi(y;x,\sigma) (x-y) \, \mathrm{d}y - f_0(x),
	\end{align}
	which are solved by the expressions for the static cargo case. 
	The terms that are first order in $v$, lead to expressions for $n_1(x)$:
	\begin{align}
	\partial_x n_0(x) &= n_1(x)(1 +c(x) p_f(x) ),\\
	\Rightarrow n_1(x) &= \frac{\partial_x n_0(x)}{1+c(x)p_f(x)},
	\end{align}
	and $f_1(x)$:
	\begin{align}
	-\partial_xf_0(x) + \frac{1}{\sigma^2} n_0(x) &= - \frac{1}{\sigma^2} c(x)n_1(x) \int_{-r}^r \phi(y;x,\sigma) (x-y) \, \mathrm{d} y -f_1(x),\\
	\Rightarrow f_1(x) &= \partial_x f_0(x) - \frac{1}{\sigma^2} n_0(x) - \frac{1}{\sigma^2} c(x) n_1(x) \int_{-r}^r \phi(y;x,\sigma) (x-y) \mathrm{d}y.
	\end{align}
	We insert this result in the general relation for the relay force (Eq.~\eqref{eq:Force}) and find that the force on a cargo moving with imposed velocity $v$ is
	\begin{align}
	\label{eq:FvSI}
	\nonumber
	F(v) &= \frac{1}{\sigma^2} \int_{-1/2}^{1/2} \mathrm{d}x \int_{-r}^r n(x,y,v) (x-y) \, \mathrm{d} y \\
	\nonumber
	& = \int_{-1/2}^{1/2} \mathrm{d}x \, f(x,v)\\
	\nonumber
	&=F_{\text{sc}} + v\int_{-1/2}^{1/2} \mathrm{d} x \, f_1(x) + \mathcal{O}(v^2) \\
	\nonumber
	&=F_{\text{sc}} +v\int_{-1/2}^{1/2} \mathrm{d} x \left(\partial_x f_0(x) - \frac{1}{\sigma^2}n_0(x) - \frac{1}{\sigma^2} c(x) \frac{\partial_x n_0(x)}{1+c(x)p_f(x)} \int_{-r}^r \phi(y;x,\sigma) (x-y) \mathrm{d}y\right) + \mathcal{O}(v^2) \\
	&=F_{\text{sc}} - \frac{v}{\sigma^2} N_{sc} - \frac{v}{\sigma^2} \int_{-1/2}^{1/2} \mathrm{d} x \,  c(x) \frac{\partial_x n_0(x)}{1+c(x)p_f(x)} \int_{-r}^r \phi(y;x,\sigma) (x-y) \mathrm{d}y + \mathcal{O}(v^2)
	\end{align}
	We use this result in the force balance equation
	$\gamma_{\text{c}} v = F(v)$ to self-consistently determine the velocity and therefore the force on a self-propelled cargo, as shown in the main text for the weak-binding limit.
	\subsection{Weak-binding limit}
	\noindent In the weak-binding limit ($c_0 \ll 1$), we can approximate the expressions for the DNA-relay force. 
	Considering only terms up to first order in $c_0$, the expression for $F_{\rm sc}$ (Eq.~\ref{eq:FscSI}) can be solved analytically:
	\begin{align}
	F_{\text{sc}} = \int_{-1/2}^{1/2} f(x) \, \mathrm{d}x 
	&= N_{\text{tot}} \frac{c_0}{\sigma^2}\int_{-1/2}^{1/2} \mathrm{d}x \, (1+mx) \int_{-r}^r \mathrm{d}y \, \phi(y;x,\sigma)(x-y) \\
	&= N_{\text{tot}} m c_0\int_{-1/2}^{1/2}  x (\phi(x;r,\sigma) - \phi(x;-r,\sigma))\, \mathrm{d}x \label{eq:FscWblIntermediateResult}\\
	&= N_{\text{tot}} c_0 m \left(\sqrt{\frac{2}{\pi}} \sigma \left(e^{-\frac{(1+2r)^2}{8\sigma^2}} - e^{-\frac{(1-2r)^2}{8\sigma^2}}\right) + r \left(\text{erf}\left(\frac{1-2r}{2\sqrt{2} \sigma} \right)+ \text{erf}\left(\frac{1+2r}{2\sqrt{2} \sigma}\right)\right)\right)
	\end{align}
	For $\sigma\ll1$, the width $\sigma$ of the Gaussian function $\phi(x;\pm r,\sigma)$ is much smaller than the system size and we find $F_{\text{sc}} \approx N_{\text{tot}} 2rc_0 m$ using Eq.~\eqref{eq:FscWblIntermediateResult}. 
	For $\sigma\gg1$ (i.e. the width of the Gaussian is much larger than the system size):
	\begin{align}
	F_{\text{sc}} &\approx N_{\text{tot}}  \frac{m c_0}{\sigma^2}\int_{-1/2}^{1/2}  x \sigma^2 \frac{1}{\sqrt{2 \pi \sigma^2}}\left(1-\frac{(x-r)^2}{2 \sigma^2}-1+\frac{(x-r)^2}{2 \sigma^2}\right)\, \mathrm{d}x\\
	&= N_{\text{tot}} 2r\frac{m c_0}{\sqrt{2 \pi \sigma^2}}\frac{1}{\sigma^2}\int_{-1/2}^{1/2}  x^2 \, \mathrm{d}x\\
	&= N_{\text{tot}} 2r\frac{m c_0}{\sqrt{2 \pi}}\frac{1}{\sigma^3} \frac{1}{12}
	\end{align}
	Therefore, we find that $F_{\text{sc}}=\text{const}$ ($\sigma \ll 1$) and $F_{\text{sc}}\propto1/\sigma^3$ ($\sigma \gg 1$), as discussed in the main text.
	The number of cargo-bound beads can be approximated in the weak-binding limit:
	\begin{equation}
	N_{\text{sc}} \approx \int_{-1/2}^{1/2} c(x) p_f(x) \, \mathrm{d}x = \int_{-1/2}^{1/2} N_{\text{tot}} c_0(1+mx) p_f(x) \, \mathrm{d}x = N_{\text{tot}} c_0 \int_{-1/2}^{1/2} p_f(x) \, \mathrm{d}x.
	\end{equation}
	In the last step we used that $p_f(x)$ is even in $x$, such that the integral over $x p_f(x)$ is zero.  
	For $\sigma$ a lot smaller than the system size, $\sigma \ll 1$, we find $N_{\text{sc}} \approx 2 r c_0 N_{\text{tot}}$.
	
	In the case of the dynamic cargo, the equation for the force Eq.~\eqref{eq:FvSI} can be approximated to linear order in $c_0$ by:
	\begin{equation}
	F(v) = F_{\text{sc}} - \frac{v}{\sigma^2} N_{\text{sc}}.
	\end{equation}	
	\subsection{Strong-binding limit}
	\begin{figure}[t!]
		\centering
		\includegraphics[width=\textwidth]{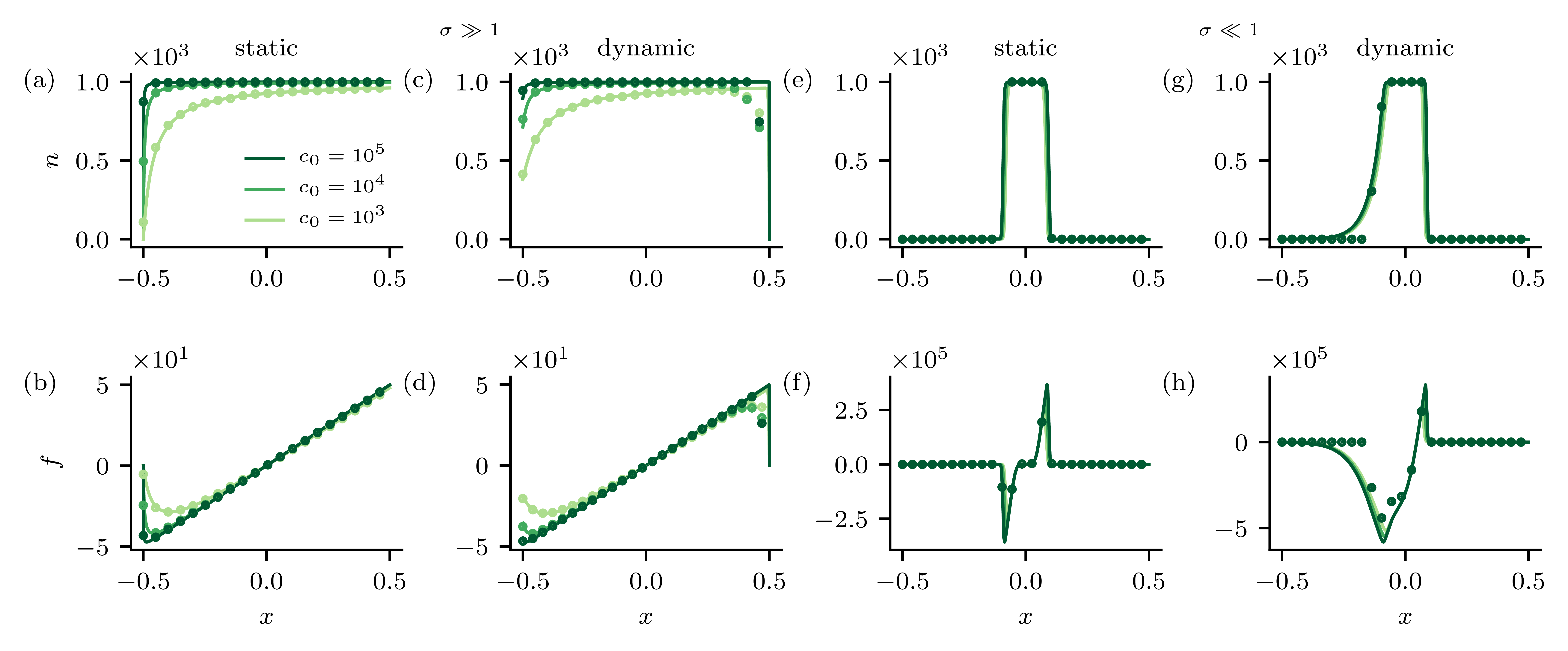}
		\caption{{\bf Distributions of the density of cargo-bound chromosomal elements $n(x)$ with rest position $x$ and the corresponding force density $f(x)$ in the strong-binding limit.} (a,b,c,d) $n(x)$ and $f(x)$ for a static ($v=0$) and a moving ($v=0.05$) cargo in the regime of large chromosomal fluctuations for different values of the binding propensity $c_0$. (e,f,g,h) $n(x)$ and $f(x)$ for a static ($v=0$) and a moving ($v=0.05$) cargo in the regime of small chromosomal fluctuations for different values of the binding propensity $c_0$. The density of cargo-bound beads $n(x)$ and the force density $f(x)$ for a static cargo are obtained from Eq.~\eqref{eq:Density} and Eq.~\eqref{eq:ForceDensity}, while the profiles for a moving cargo are obtained from a numerical solution of Eq.~\eqref{eq:nxt} and Eq.~\eqref{eq:fxt}. We compare results from simulations (dots) and theory (lines).}
		\label{fig:Distributions_sbl}
	\end{figure}
\begin{figure}[t!]
		\centering
		\includegraphics[width=\textwidth]{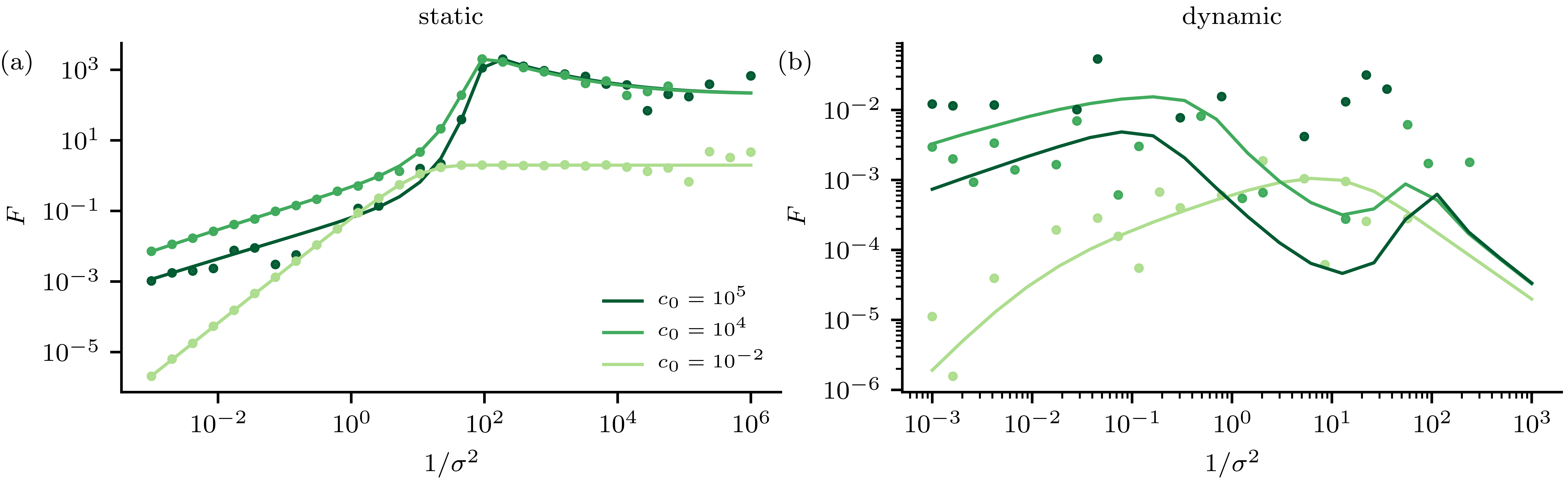}
		\caption{{\bf Average relay force $F$ in the strong-binding limit ($c_0\gg1$) for different values of the binding propensity $c_0$ and different values of the bead fluctuations $\sigma$.} We compare results from simulations (dots) with our theory (lines). (a) Force on a static cargo in the strong-binding limit. (b) Force on a dynamic cargo in the strong-binding limit.}
		\label{fig:DynamicCargoForceStrongBinding}
	\end{figure}

	
	\noindent While in the weak-binding limit saturation effects can be neglected, we now explain how the relay force is altered in the strong-binding limit. In analogy to our discussion in the main text, we first consider the static cargo case.	To understand the effect of an increase of $c_0$ on force generation, we consider the density profiles of cargo-bound chromosomal elements $n(x,t)$ with rest position $x$ (Fig.~\ref{fig:Distributions_sbl}a,e). In the case of large chromosome fluctuations ($\sigma\gg1$), beads can attach to the cargo from distances on the order of the system size and hence finite size effects matter. An increase in $c_0$ leads to a binding profile $n(x,t)$ that quickly approaches a uniform distribution (Fig.~\ref{fig:Distributions_sbl}a). When all beads bind to the cargo with the same probability, no net force is generated. This observation explains the reduction of the force for large bead fluctuations for very strong binding (Fig.~\ref{fig:DynamicCargoForceStrongBinding}a). On the other hand, for small chromosome fluctuations ($\sigma\ll1$), the chromosomal region from which beads attach to the cargo is smaller than the system size. In the strong-binding limit, beads with a rest position below the cargo are nearly always bound to the cargo (Fig.~\ref{fig:Distributions_sbl}e). However, there is a region in front and behind of the cargo, for which the binding density $n(x,t)$ is not saturated ($0<n<1$). Due to the gradient in ParA, more beads bind to the cargo from ahead than behind of the cargo, resulting in a positive net force. Increasing $c_0$ further, enlarges the region around the cargo where $n(x,t)$ is saturated, but as long as this region is smaller than the system size, there is an imbalance of the cargo-bound beads and hence a net force is generated (Fig.~\ref{fig:DynamicCargoForceStrongBinding}a).
	
	In the case of a dynamic cargo, the generated force is the result of the driving force and additional friction due to cargo-bound beads. For very large binding propensities, not only the driving force is reduced (for large fluctuations) as discussed above, but also the number of cargo-bound beads is increased (Fig.~\ref{fig:Distributions_sbl}c,g), which leads to larger effective friction coefficients for the cargo (Fig.~\ref{fig:Distributions_sbl}d,h).
	
\bibliography{references.bib}